\documentclass[fleqn,12pt,twoside]{article}
\usepackage{espcrc1}

\usepackage{graphicx}
\usepackage{rotating}

\newcommand{\AmS}{{\protect\the\textfont2
  A\kern-.1667em\lower.5ex\hbox{M}\kern-.125emS}}

\title{Monte Carlo Models: {\it Quo Vadimus}?}

\author{Xin-Nian Wang\address{Nuclear Science Division, MS 70-319,
    Lawrence Berkeley National Laboratory, Berkeley, CA 94547}%
        \thanks{This work was supported by the Director, Office of Energy
          Research, Office of High Energy and Nuclear Physics, Divisions of 
          Nuclear Physics, of the U.S. Department of Energy under 
          Contract No.\ DE-AC03-76SF00098}}
       
\begin{document}

\maketitle

\begin{abstract}
Coherence, multiple scattering and the interplay between soft 
and hard processes are discussed. These physics phenomena are 
essential for understanding the nuclear dependences of rapidity
density and $p_T$ spectra in high-energy heavy-ion collisions. 
The RHIC data have shown the onset of hard processes and
indications of high $p_T$ spectra suppression due to 
parton energy loss. Within the pQCD parton model, the combination 
of azimuthal anisotropy ($v_2$) and hadron spectra suppression 
at large $p_T$ can help one to determine the initial gluon density 
in heavy-ion collisions at RHIC.
\end{abstract}

\textfloatsep=0.2in

\section{Introduction}

Monte Carlo models have played an important and irreplaceable role in both
low and high-energy physics. First of all, they are needed to simulate 
the physics and detector capabilities for any proposed experiment. They 
are necessary to study the efficiency, acceptance corrections and 
background in the analyses of experimental data. They are also 
very useful tools for theorists to test and estimate the consequences 
of new physics ideas and phenomena. This is especially true for high-energy
heavy-ion collisions. 

In order to study the properties of the quark-gluon plasma (QGP) produced 
in the early stage of heavy-ion collisions and to search for evidence of 
the QCD phase transition, one needs to understand the whole evolution
history of the collisions. Since the reaction dynamics is very complex 
and there is not a simple standard analytic model, one has to rely on 
Monte Carlo models to incorporate many aspects of strong interactions
in the simulation of heavy-ion collisions.
The models and parameters therein can be constrained by well-tested 
theories and experimental data especially in $pp$ and $pA$ collisions
where QGP is not expected to form. They provide a baseline calculation
of physical observables in the absence of new physics due to the 
formation of QGP. However, large uncertainties exist
in extrapolating to $AA$ collisions. Comparisons of results produced
by different Monte Carlo models or by varying the model parameters
are useful to provide a measure of the extrapolation uncertainties.
The latest RHIC data \cite{brahms,phenix,phobos,star} 
have proven to provide good
constraints on Monte Carlo models that in turn help us to understand
the reaction dynamics and the initial conditions. Finally, Monte
Carlo models also serve as theoretical laboratories to test proposed
signals and probes of QGP such as jet quenching \cite{Wang:1992xy}.

There are many Monte Carlo models for heavy-ion collisions, reflecting 
the complexity of the problems. But in general 
they can be divided into three categories. Hadron and string based 
models, {\it e.g.}, FRITIOF \cite{fritiof}, DPM \cite{dpm}, 
VENUS \cite{venus}, RQMD \cite{rqmd}, ARC \cite{arc}, ART \cite{art},
URQMD \cite{urqmd}, LUCIAE \cite{luciae} and others, consider hadrons
and Lund strings \cite{lund} as the effective degrees of freedom in 
strong interactions. Depending on the center-of-mass energy of a 
two-body collision, particles are produced through resonances 
and strings which hadronize according to the Lund string fragmentation
model \cite{lund}. These models are adequate for
particle production and rescattering in hadronic and
nuclear collisions at energies around or below the CERN-SPS,
in which partonic degrees of freedom are not yet important.
At high energies ($\sqrt{s}> 50$ GeV), hard or semi-hard partonic
scatterings become important and contribute significantly to the
particle production and dominate the high $p_T$
hadron spectra even at the CERN-SPS energy. In this case, pQCD-inspired 
models, {\it e.g.}, HIJING \cite{hijing}, VNI \cite{vni}
NEXUS \cite{nexus} and AMPT \cite{ampt}, are more relevant. 
The third class of models, such as LEXUS \cite{lexus}, 
pQCD parton model \cite{xnw00,levai}, parton cascade models \cite{pcm}
and hydrodynamic models \cite{hydro}, are not strictly event 
generators. But they are useful to help us to understand particle 
production and evolution in heavy-ion collisions, many aspects of which 
are difficult to simulate in an event generator. 

I will not give a review of these models \cite{oscar} here. Instead,
I will concentrate on the parton based models and discuss the essential
physics behind them. In particular, I will discuss the interplay between 
soft and hard processes and their different behaviors in multiple 
parton scattering processes. I will demonstrate the difficulties
in incorporating quantum interferences in Monte Carlo models and
how one can use mission specific models such as the pQCD parton model to
overcome these difficulties.

\section{Soft versus hard processes}
 
Not long after the importance of minijets in both high-energy 
$pp(\bar p)$ \cite{Wang:1989bw} and heavy-ion collisions \cite{minijet}
was realized, two parton-based Monte Carlo models \cite{hijing,vni} were 
developed to incorporate hard processes in high-energy 
heavy-ion collisions. The jet production cross section in these hard 
processes can be described very well by the pQCD parton model when the 
transverse momentum transfer $p_T$ involved is very large. However, 
the pQCD parton model calculation of jet cross sections diverges when 
$p_T\rightarrow 0$ and the processes become non-perturbative.
No theoretical model exists that can describe both the large and small
$p_T$ processes. In order to combine hard processes and
non-perturbative soft processes in a Monte Carlo model, a
cut-off scale $p_0$ in transverse momentum transfer of parton
scattering is introduced that phenomenologically separates soft 
and hard processes. Assuming eikonalization of these processes, one 
can calculate the $pp(\bar p)$ cross section in this two-component
model \cite{hijing},
\begin{equation}
\sigma_{\rm in}^{NN}=\int d^2b \left[1-
e^{-(\sigma_{\rm jet}+\sigma_{\rm soft})T_{NN}(b)}\right],
\end{equation}
where $T_{NN}(b)$ is the nucleon-nucleon overlap function,
$\sigma_{\rm soft}$ is a parameter to represent inclusive soft
parton cross section and $\sigma_{\rm jet}$ is calculated from
pQCD parton model,
\begin{equation}
\sigma_{\rm jet}=\int_{p_0^2}^{s/4} dp^2_T dy_1 dy_2
\frac{1}{2}\sum_{a,b,c,d} x_1 x_2 f_a(x_1)f_b(x_2) 
\frac{d\sigma_{ab\rightarrow cd}}{d\hat{t}}.
\label{eq:jet}
\end{equation}
The two parameters, $p_0$ and $\sigma_{\rm jet}$ in this two-component
model can be fixed by the experimental data on the cross sections 
of $pp(\bar p)$ collisions and the corresponding $dN_{\rm ch}/d\eta$.
This is the basic principle behind HIJING \cite{hijing}, 
VNI \cite{vni} and the recent NEXUS \cite{nexus}. In this two-component
model, the rapidity density of charged multiplicity can
be expressed as
\begin{equation}
\frac{dN_{\rm ch}^{NN}}{d\eta}=\langle n\rangle_{\rm soft}
+\langle n\rangle_{\rm hard} 
\frac{\sigma_{\rm jet}^{NN}(s)}{\sigma_{\rm in}^{NN}(s)}.
\end{equation}
In HIJING, soft particle production is modeled by a string model 
whose contribution to $dN_{\rm ch}/d\eta$ is almost constant. 
The contribution from minijet production, on the other hand, 
increases with energy due to the increase of the inclusive jet cross section.

To extend to nuclear collisions, one has to consider the nuclear 
dependences of these two components. For the soft part, the interaction
is coherent and its contribution to the multiplicity should be 
proportional to the number of wounded nucleons. On the other hand,
minijet production is approximately incoherent and proportional
to the number of binary collisions, modulo the nuclear shadowing
of parton distributions and jet quenching. The total charged
hadronic rapidity density will acquire the form
\begin{equation}
\frac{1}{\langle N_{\rm part}\rangle/2}\frac{dN_{ch}^{AA}}{d\eta}
=\langle n\rangle_{\rm soft} + \langle n\rangle_{\rm hard} 
\frac{\langle N_{\rm binary}\rangle}{\langle N_{\rm part}\rangle/2}
\frac{\sigma_{\rm jet}^{AA}(s)}{\sigma_{\rm in}^{NN}},
 \label{eq:nch} 
\end{equation}
where $\sigma_{\rm jet}^{AA}(s)$ is the averaged inclusive jet cross 
section per $NN$ collision in $AA$, which is generally smaller 
than $\sigma_{\rm jet}^{NN}$ due to nuclear shadowing of parton
distributions. The energy dependence of the minijet cross section 
$\sigma_{\rm jet}^{AA}(s)$ in the hadronic multiplicity is then 
amplified by 
$\langle N_{\rm binary}\rangle/\langle N_{\rm part}\rangle \sim A^{1/3}$ 
in heavy-ion collisions relative to $pp(\bar{p})$ by the binary nature of 
semi-hard processes.

\begin{figure}[htb]
\begin{minipage}[t]{78mm}
\includegraphics[scale=0.53]{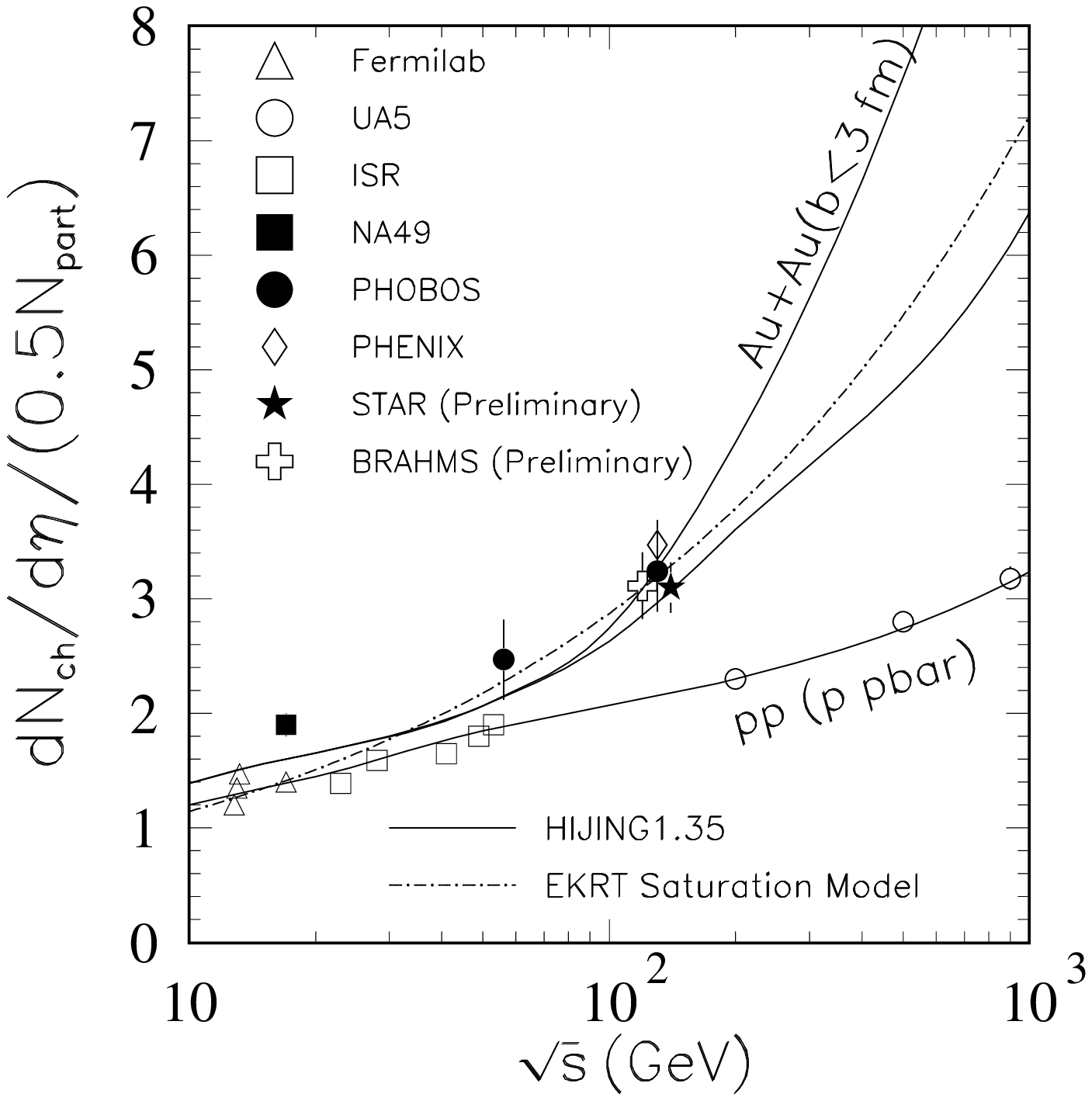}
\vspace{-0.3in}
\caption{The RHIC data\protect\cite{brahms,phenix,phobos,star} 
for  central Au+Au are  compared to $pp$ and $p\bar{p}$ 
data \protect\cite{ppdata} and the NA49 $Pb+Pb$ data \protect\cite{na49}.
HIJING1.35 (solid) with (upper) and without jet quenching (lower) 
and EKRT (dot-dashed) predictions \protect\cite{ekrt} are also shown.}
\label{fig:dndeta0}
\end{minipage}
\hspace{\fill}
\begin{minipage}[t]{78mm}
\includegraphics[scale=0.53]{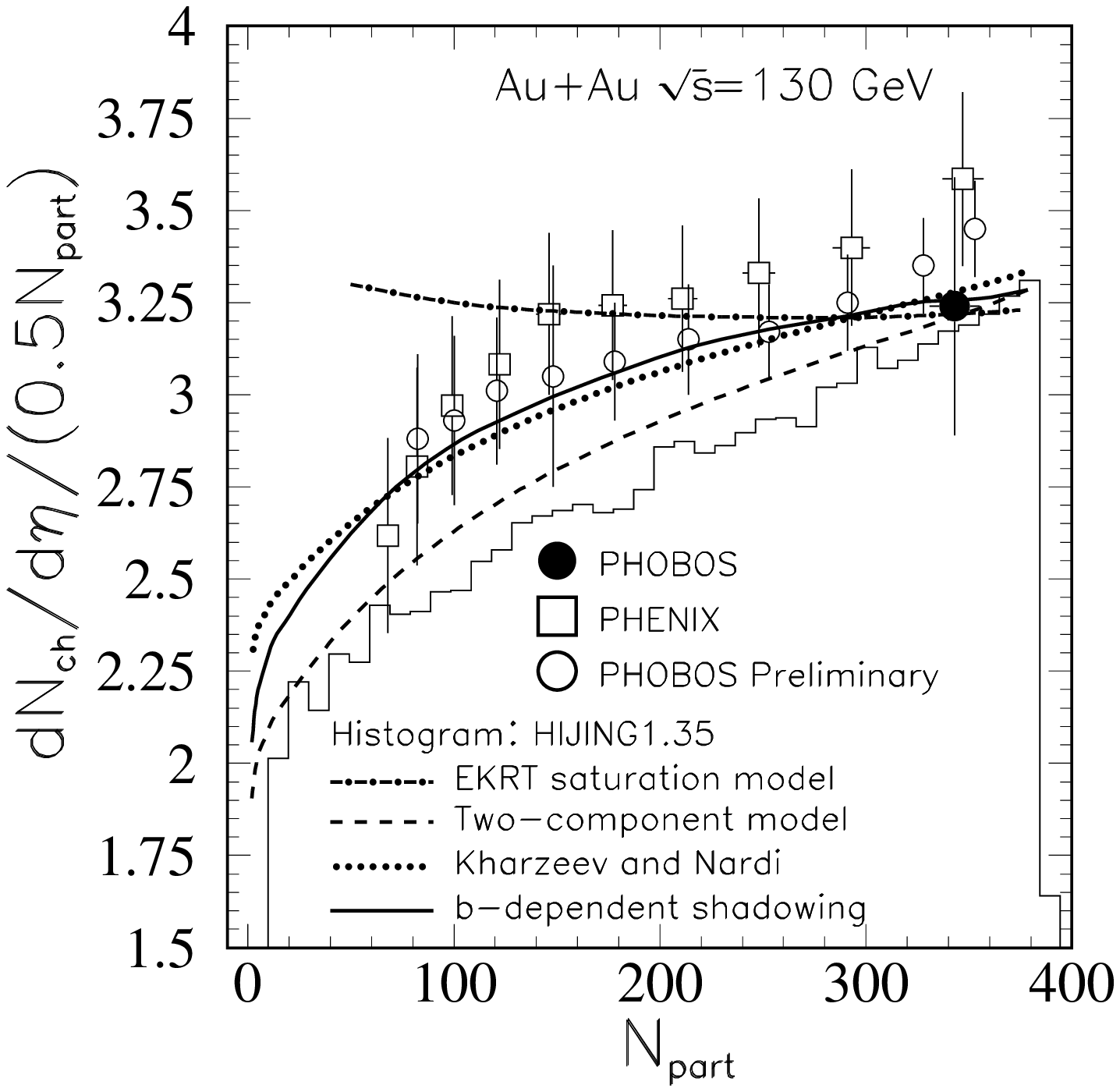}
\vspace{-0.3in}
\caption{The RHIC data \cite{phenix,phobos} on centrality dependence 
of $dN_{\rm ch}/d\eta/0.5\langle N_{\rm part}\rangle$ are compared with
HIJING and EKRT \cite{ekrt} predictions. The two-component model in 
Eq.~(\protect\ref{eq:nch}) with (solid)  and without (dashed) 
$b$-dependence of shadowing are also shown.}
\label{fig:central}
\end{minipage}
\end{figure}

Shown in Fig.~\ref{fig:dndeta0} are HIJING results of $dN_{ch}/d\eta$ per
pair of participant nucleons in $pp(p\bar{p})$ and central $Au+Au$ 
collisions at different colliding energies as compared to experimental
data. The rise of the multiplicity per participant relative to SPS is 
consistent with the 
predicted binary scaling of the hard component from $pp$ to $AA$ via  
Eq.~(\ref{eq:nch}) without jet quenching. Final state interaction such 
as jet quenching will enhance the $A$ dependence of $dN_{ch}/d\eta$ 
especially at high energies.

Also shown in Fig.~\ref{fig:dndeta0} is the result of a final-state 
saturation model by EKRT \cite{ekrt}. In this model, the pQCD 
growth of low $p_T$ gluons is cut off below a saturation scale
which is determined by the local saturation requirement, 
$dN_g/dyd^2s=p_{sat}^2/\pi$. Here $dN_g/dyd^2s$ is the gluon rapidity
density per unit transverse area. The resultant saturation scale 
increases with energy and $\langle N_{\rm part}\rangle$. Assuming direct 
proportionality between parton and the final hadron number, the 
energy and nuclear dependence of the total hadronic  
rapidity density in central $A+A$ collisions can then be estimated.
As shown, the EKRT result is remarkably close to the HIJING results 
and is also consistent with the RHIC data.

However, a critical difference between these models
is the centrality dependence of $dN_{ch}/d\eta\langle N_{\rm part}\rangle$.
In the HIJING two-component model of minijet production the multiplicity per 
participant increases with $A$ according to Eq.(\ref{eq:nch}) while
it actually {\em decreases} in the EKRT model with increasing $A$
due to the saturation requirement. In order to distinguish the fixed-scale 
and saturation models of entropy production, the study of 
centrality dependence of the hadronic multiplicity per participant
was proposed \cite{central}.

Shown in Fig.~\ref{fig:central} are the $dN_{ch}/d\eta$ per participant 
pair as functions of  $\langle N_{\rm part}\rangle$.
The EKRT model \cite{ekrt} has a weakly decreasing or constant dependence 
on centrality in semi-peripheral to central collisions. The HIJING 
result, on the other hand, increases monotonically with the number 
of binary collisions per participant 
$\langle N_{\rm binary}\rangle/\langle N_{\rm part}\rangle$ 
in an intuitive way as given in Eq.(\ref{eq:nch}).
The RHIC data clearly favor the the latter case.
Naively, $\langle N_{\rm binary}\rangle/\langle N_{\rm part}\rangle 
\sim \langle N_{\rm part}\rangle^{1/3}$. The deviation from such a simple 
dependence in HIJING calculation might be due to a combined effect of 
jet quenching and dilute edges in nuclear distributions used in HIJING.
Also shown as dashed line is the simple two-component model in
Eq.~(\ref{eq:nch}) which agrees with the HIJING result.
The parameters, $\langle n\rangle_{\rm soft}\approx 1.6$ 
and $\langle n\rangle_{\rm hard}\approx 2.2$ are determined from
a fit to the measured energy dependence of $dn_{\rm ch}/d\eta$ in
$pp(\bar p)$ collisions. Another phenomenological two-component model by 
Kharzeev and Nardi \cite{kharzeev}, shown as the dotted line, assumed
$dn^{pp}_{\rm ch}/d\eta=2.25$ when fixing the parameters.
It coincides exactly with the dashed line if $dn^{pp}_{\rm ch}/d\eta=2.0$
is used.

The jet cross section $\sigma_{\rm jet}^{AA}$ used in the two-component
model (dashed line) is calculated with nuclear shadowing of parton 
distributions averaged over mini-biased events. Shadowing reduces
the effective jet cross section and thus minijet contribution to charged
multiplicity. Without shadowing, the two-component model will over-predict 
the charged multiplicity in central $Au+Au$ collisions. However, parton
shadowing should depend on the impact-parameter of nuclear collisions.
If an impact-parameter dependent parton shadowing as described
in HIJING \cite{hijing} is used, the effective jet cross section 
$\sigma_{\rm jet}^{AA}$ should also depend on centrality. Using this
$b$-dependent cross section in Eq.~(\ref{eq:nch}), one
gets the solid line in Fig.~\ref{fig:central}. The RHIC data certainly
favor this scenario. Such shadowing effect is similar to the initial
state saturation \cite{mclerran}. A phenomenological implementation
of the initial state saturation is shown \cite{kharzeev} to give 
a similar result as the two-component model. Another independent constraint
is the rapidity dependence of $dN_{ch}/d\eta/\langle N_{part}\rangle$.
PHOBOS preliminary \cite{phobos} data are consistent with HIJING model.

\section{Interplay between soft and hard processes}

The centrality dependences of soft and hard processes not only manifest
themselves in the total charged multiplicity but also in the hadron $p_T$ 
spectra. Incoherent hard parton scattering should dominate the hadron
spectra at high $p_T$ while coherent soft interactions contribute
mainly to the low $p_T$ region. This leads to some nontrivial
nuclear dependence of hadron $p_T$ spectra. A good Monte Carlo
model should be able to simulate this nuclear dependence.

In a schematic Glauber model of multiple parton scattering, one can
calculate the produced parton spectra \cite{ewxw} in $p+A$ collisions:
\begin{equation}
        \frac{E}{A}\frac{d\sigma^j_{NA}}{d^3p} \approx 
        E\frac {d\sigma_{NN}^j}{d^3p} +
        \frac {9A^{1/3}}{16\pi r^2_0} \sum_i \int 
        \frac {dx_i}{x_i} f_{i/N}(x_i)
       \left[ \sum_k\int\frac{d^3p_k}{E_k}
        h^k_{iN}h^j_{kN} - (\sigma_{iN}+\sigma_{jN})h^j_{iN}\right],
      \label{eq:ra}
\end{equation}
where $h^{j}_{iN}$ is the differential cross
section for parton-nucleon scattering $i+N \to j+X$, and $f_{i/N}(x_i)$
is the parton distribution in a nucleon.
The effective parton-nucleon total cross section is defined as
$\sigma_{iN}(p_i)=\frac{1}{2} \sum_j \int \frac {d^3p_j}{E_j}
h^{j}_{iN}(p_i,p_j)$
and the differential nucleon-nucleon cross section for parton 
production as
$ Ed\sigma_{NN}^j/d^3p
=\sum_i \int \frac {dx_i}{x_i} f_{i/N}(x_i) h^j_{iN}(p_i,p)$.
We have taken a nucleus to be a hard sphere of radius
$r_A=r_0A^{1/3}$.

\begin{figure}[htb]
\begin{minipage}[t]{78mm}
\includegraphics[scale=0.53]{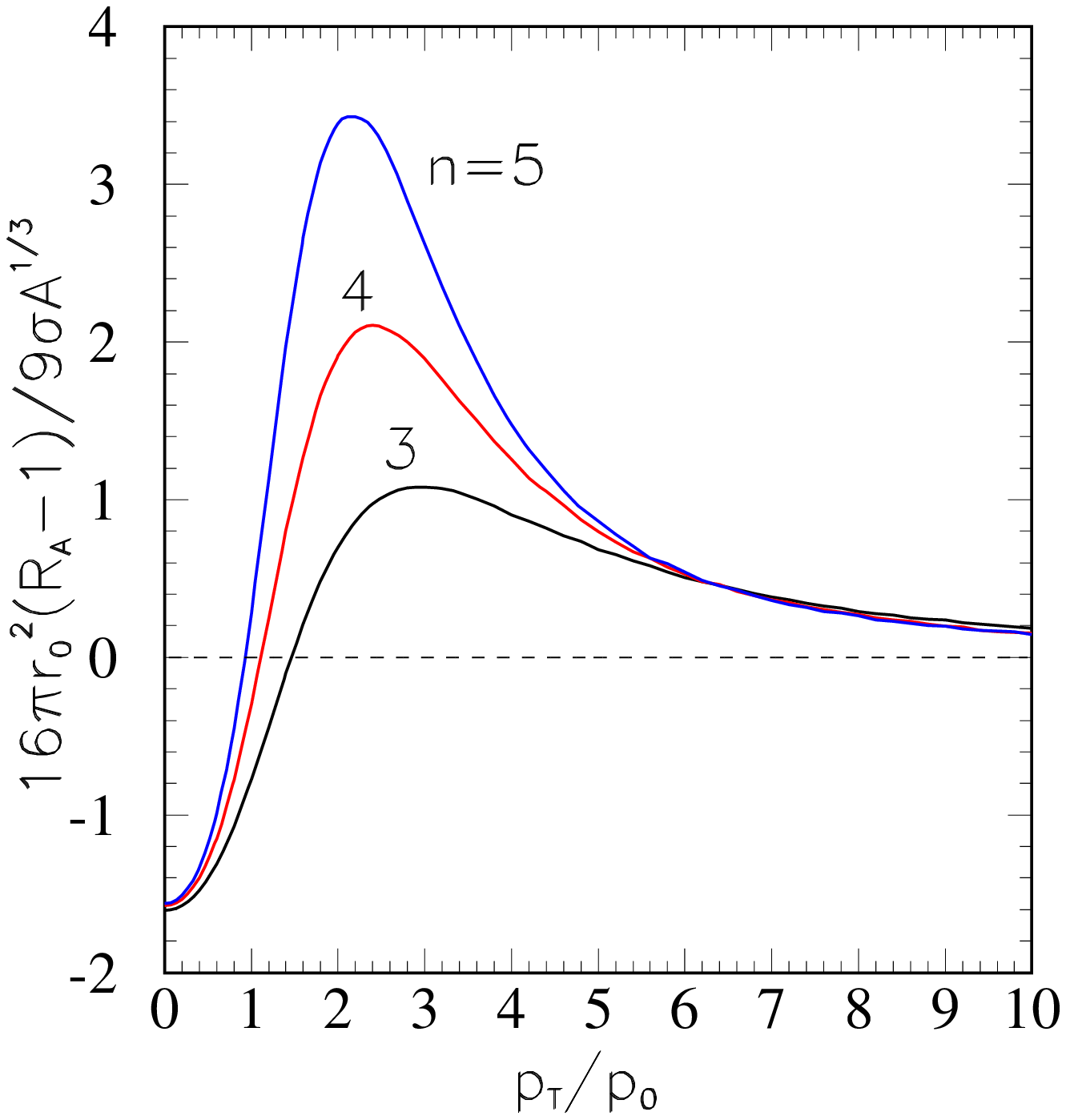}
\vspace{-0.3in}
\caption{The nuclear modification factor in a schematic model of 
multiple parton scattering with a simple form of 
parton-nucleon cross section $1/(p_T^2+p_0^2)^n$.}
\label{fig:schem}
\end{minipage}
\hspace{\fill}
\begin{minipage}[t]{78mm}
\includegraphics[width=2.9in,height=2.9in]{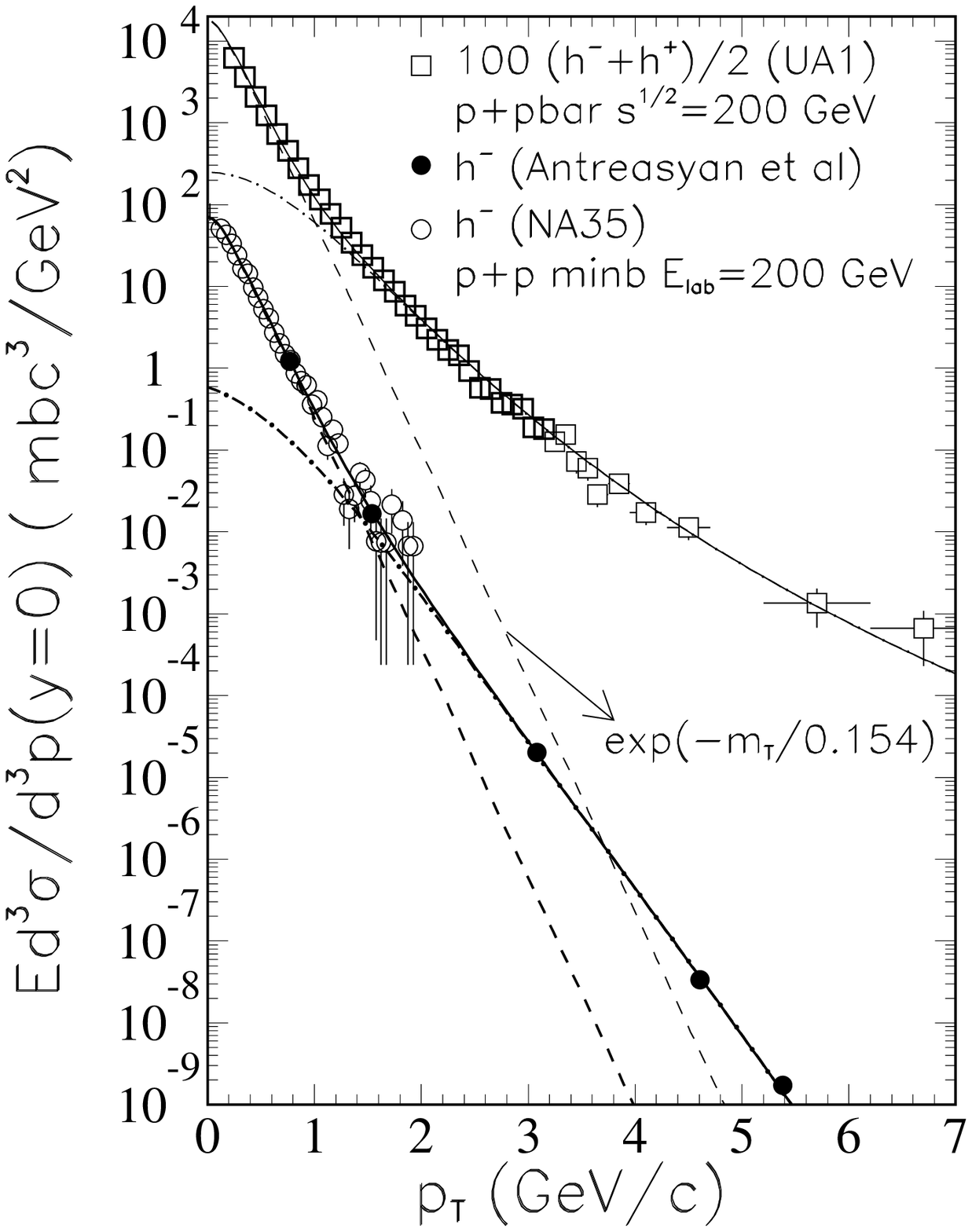}
\vspace{-0.3in}
\caption{Experimental data \cite{ppdata} on hadron spectra 
in $pp(\bar p)$ collisions are compared to two-component 
(power-law and exponential) parameterization.}
\label{fig:para}
\end{minipage}
\end{figure}

\begin{figure}[htb]
\begin{minipage}[t]{78mm}
\includegraphics[scale=0.43]{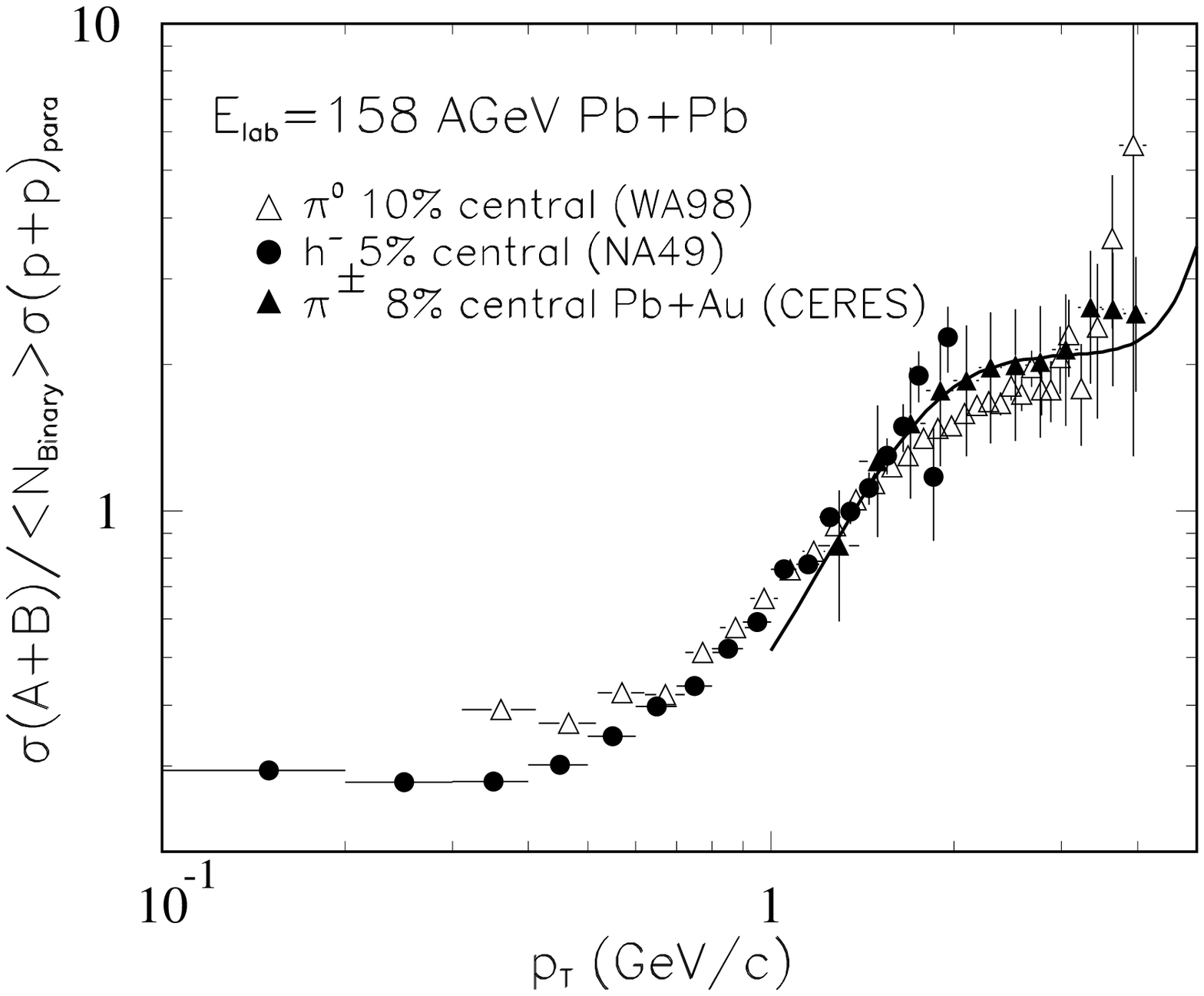}
\vspace{-0.3in}
\caption{The nuclear modification factor for hadron spectra in central
$Pb+Pb$ collisions at the CERN-SPS. The solid line is pQCD parton
model calculation}
\label{fig:r-sps}
\end{minipage}
\hspace{\fill}
\begin{minipage}[t]{78mm}
\includegraphics[scale=0.43]{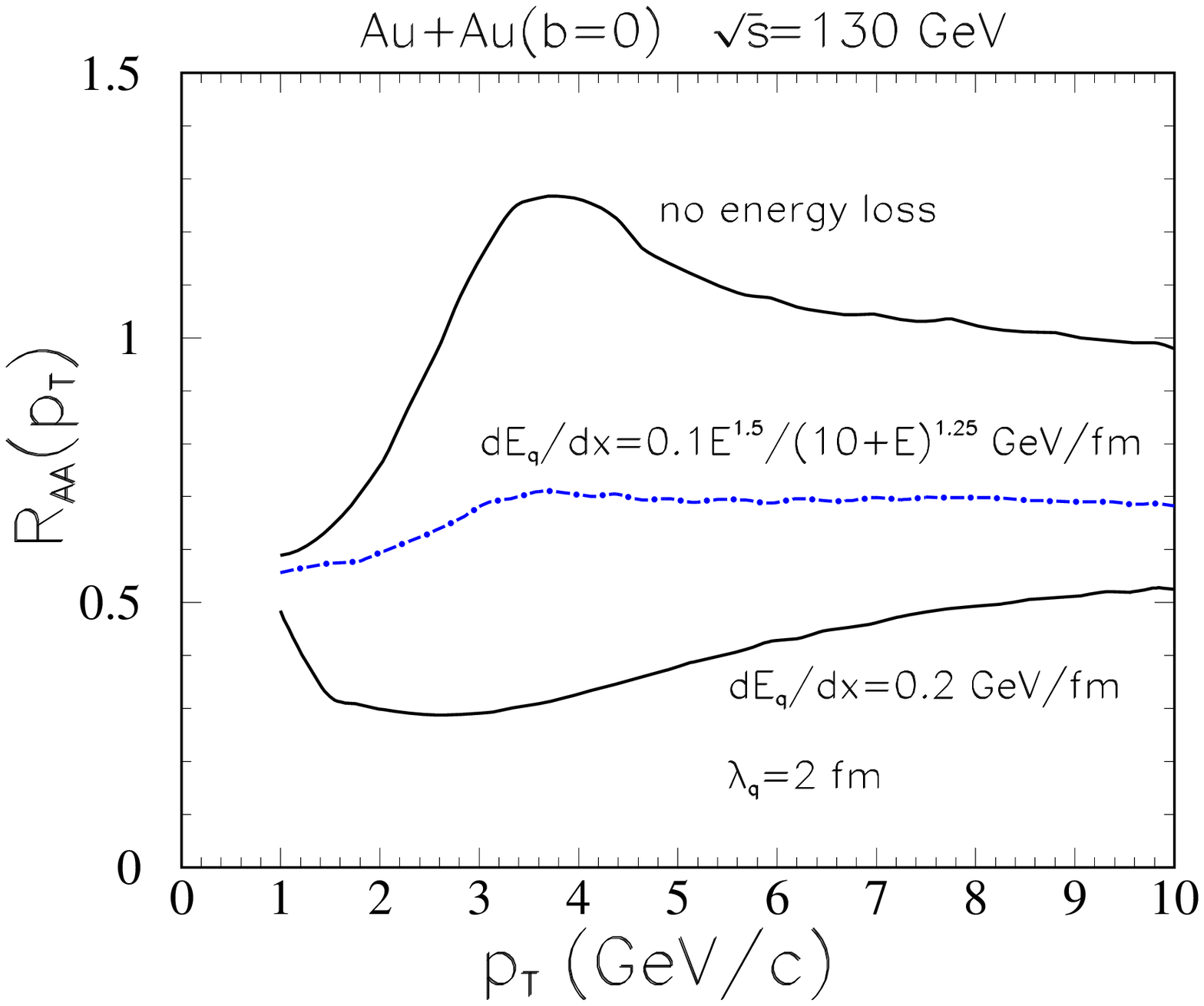}
\vspace{-0.3in}
\caption{The nuclear modification factor for hadron spectra in central
$Au+Au$ collisions at RHIC with different values of parton energy loss.}
\label{fig:r-rhic}
\end{minipage}
\end{figure}

In the above equation, the first term corresponds to the incoherent
sum of single parton scatterings in a nucleus.
The second term gives the nuclear
modification of the parton spectrum due to multiple parton scattering
inside a nucleus. This term contains contributions from both the
double parton scattering and the negative absorptive correction.
For a schematic study, let us assume that all partons are identical and
the differential parton-nucleon cross sections have a simple
regularized power-law form $ h^j_{iN}\equiv h(p_T)=C/(p_T^2+p_0^2)^n$. 
The parameter $p_0$ can be considered as the scale separating soft 
and hard processes. One can then calculate the nuclear modification
factor of the parton spectra,
\begin{equation}
R_A \equiv \frac{Ed\sigma_{NA}/d^3p}{AEd\sigma_{NN}/d^3p},
\end{equation}
which is shown in Fig.~\ref{fig:schem} for three different
values of $n$ as a function of $p_T/p_0$.
Relative to the additive model of incoherent hard scattering, 
the spectra are enhanced at large $p_T$ ($R_A>1$) for hard processes due to
multiple scattering. For soft processes at low $p_T$, the absorptive 
correction is large and the second term in Eq.~(\ref{eq:ra}) becomes 
negative, leading to the suppression of hadron spectra ($R_A<1$).
The transition between suppression and enhancement occurs at around 
the scale $p_0$ that separates soft and hard processes. If we analyze
the hadron spectra in $pp(\bar p)$ collisions as shown in 
Fig.~\ref{fig:para}, one can indeed see the two underlying components:
one in power-law form that dominates spectra at high $p_T$ and another
in exponential form for low $p_T$ hadrons. 
The transition between
these two components occurs at around $p_0=1.5-2$ GeV/$c$. 
Analyses of the hadron spectra in $pA$ collisions \cite{xnw00,ewxw} 
in terms of the  nuclear modification factor indeed show such a feature 
in $R_A(p_T)$ and the transition scale is about $p_0\approx 1.5$ GeV/$c$.
These same features from initial multiple parton scattering should also 
appear in heavy-ion collisions if there is no final state interaction.
Shown in Fig.~\ref{fig:r-sps} are the experimental data of central
$Pb+Pb$ collisions at the CERN-SPS. 
These data have almost exactly the same behavior as in $pA$
collisions at the same energy. The soft-hard transition scale is also
around $p_0\approx 1.5$ GeV/$c$. However, there seems to be no evidence
of jet quenching at SPS \cite{xnw00} which would suppress the high $p_T$ 
spectra.

As demonstrated in the schematic model, the unique features of the nuclear
dependence of the hadron $p_T$ spectra are consequences of
the intricate quantum interference between multiple parton scattering
and absorptive processes. At low $p_T$, the absorptive processes are
dominant and they reduce the contribution of independent parton scattering.
Consequently, the hadron production at low $p_T$ appears to be coherent.
The inclusive differential cross sections are only proportional
to the surface area of the colliding nuclei or the number of participant
nucleons. At large $p_T$, the destructive interference between multiple
parton scattering and absorptive processes is almost complete. Hadron
production processes at large $p_T$ become incoherent. The corresponding
inclusive cross sections are proportional to the volume of the colliding
nuclei or the number of binary nucleon-nucleon collisions. It is difficult
to simulate this quantum interference in a Monte Carlo model, especially
in the transition region where both processes are important. In HIJING,
the coherent soft processes are modeled by Lund-like strings whose number
is proportional to the number of participant nucleons.
The hard processes
are treated as incoherent parton scatterings (modulo nuclear shadowing
of parton distributions) that are proportional to the number of binary
collisions. Because of parton fragmentation, hard processes still
contribute to hadron production at low $p_T$. However, since a $p_T$ 
cut-off at $p_0\sim 2$ GeV/$c$ is imposed in HIJING, hard contributions
to hadron production become smaller at lower $p_T$. Soft processes
also contribute to hadron production with an exponential-like spectrum. 
Therefore, the resultant hadron spectra in HIJING have a smooth
transition from soft to hard processes as $p_T$ is 
increased \cite{Wang:1992xy}.
However, the string fragmentation used for jets tend to suppress the
$p_T$ broadening due to multiple parton scattering. So HIJING cannot 
reproduce the Cronin enhancement of hadron spectra at intermediate $p_T$.

\section{Parton model and high $p_T$ spectra}

The transition from soft to hard processes is only one example of
quantum interferences one has to consider in a Monte Carlo model
for high-energy heavy-ion collisions. Other effects include LPM
interference effect for a propagation of an energetic parton inside
a medium, coherent multiple scattering in a dense medium, and effects
of chiral symmetry restoration. Implementing these phenomena correctly
in a Monte Carlo model that can simulate every stage of heavy-ion 
collisions is very difficult, if not impossible. We can, however, 
resort to some mission-specific models to help us to understand
the physics of some specific aspect of heavy-ion collisions.
For example, hydrodynamic models \cite{hydro} can help us to 
understand the collective phenomena in situations where cascade 
models become invalid because of coherent scattering in a dense medium. 
Parton cascade models \cite{pcm} can assist us in understanding the 
process of parton thermalization. In the following, I will discuss the
pQCD parton model \cite{xnw00,levai}, which is very useful for the
study of high $p_T$ spectra from jets. They can be used as a probe
of the dense partonic matter in heavy-ion collisions.

\begin{figure}[htb]
\begin{minipage}[t]{78mm}
\includegraphics[scale=0.53]{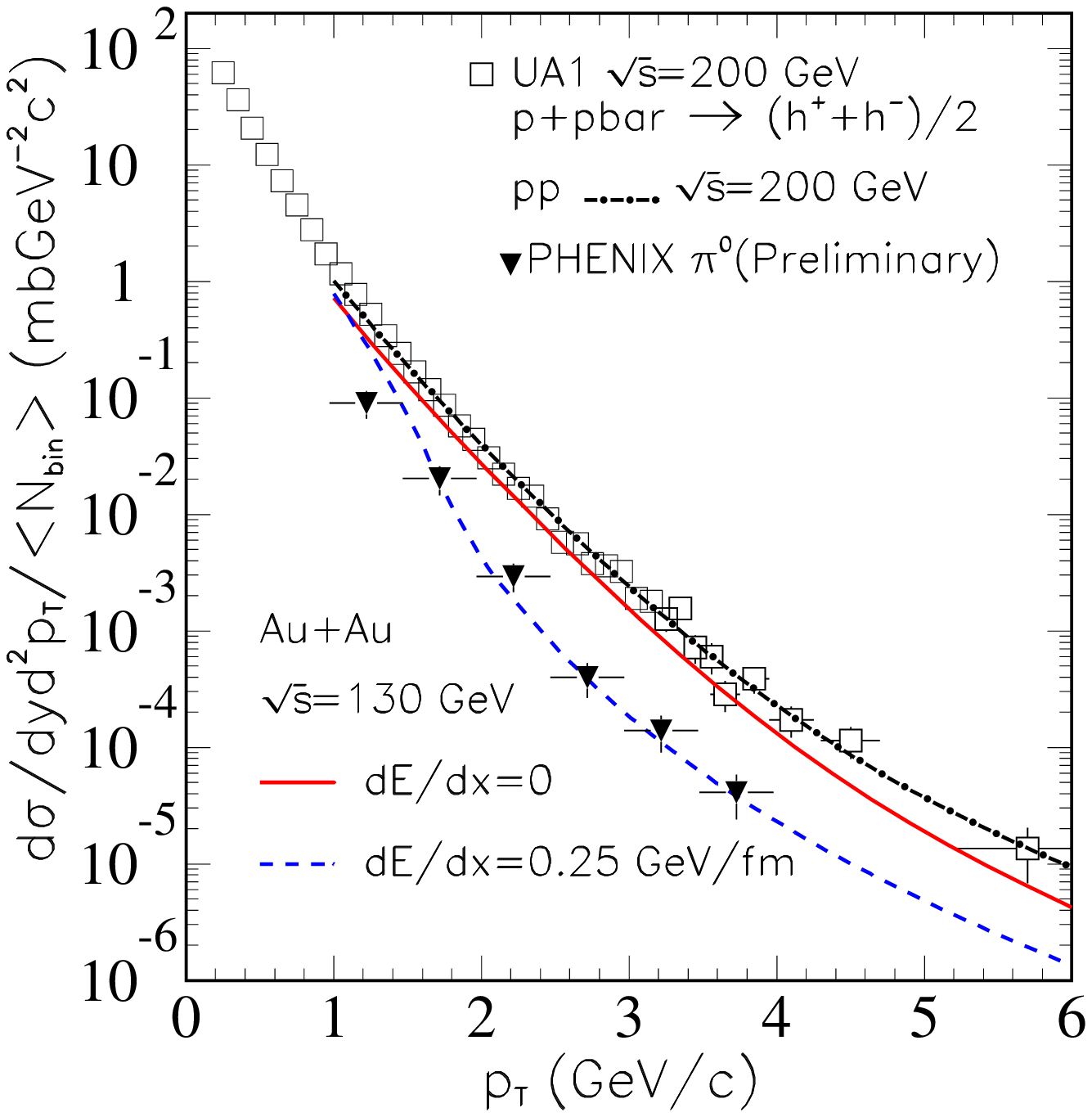}

\vspace{-0.3in}
\caption{The preliminary $\pi_0$ spectra in central $Au+Au$
at RHIC \cite{phenix} as compared to $p\bar p$ 
data \protect\cite{ppdata} and parton model calculations.}
\label{fig:pi0}
\end{minipage}
\hspace{\fill}
\begin{minipage}[t]{78mm}
\includegraphics[width=3.2in,height=3.2in]{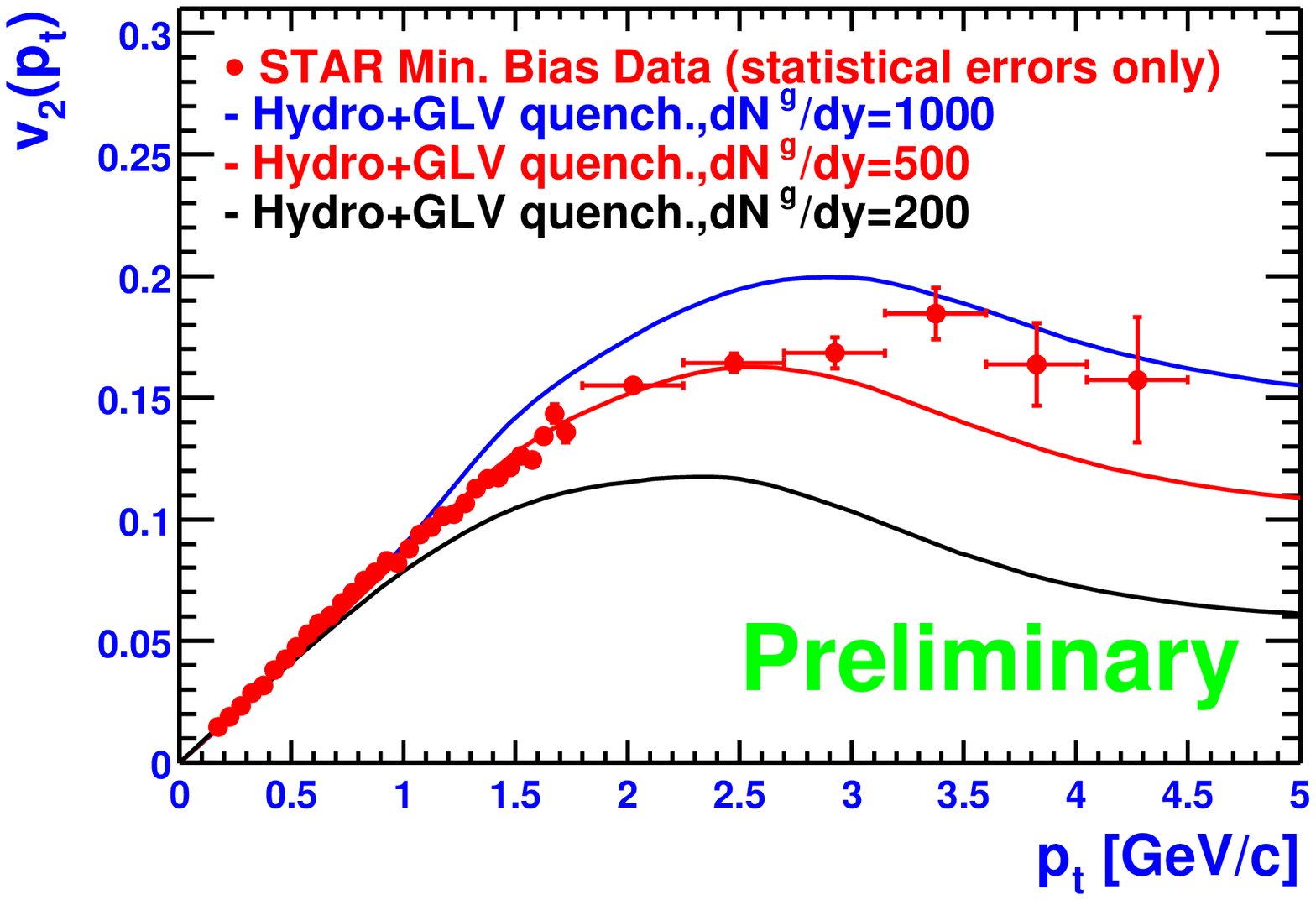}
\vspace{-0.5in}
\caption{The hydro+parton model calculation of $v_2(p_t)$ as compared
to preliminary STAR data \protect\cite{star}.}
\label{fig:v2}
\end{minipage}
\end{figure}
For high $p_T$ hadron spectra, one can essentially neglect the soft
processes and only consider hadron production from hadronization
of jets whose cross section can be described by Eq.~(\ref{eq:jet}).
Hadron spectra in $pp$, $pA$ collisions can be described 
well \cite{xnw00} within this parton model. Constrained by the 
existing $pA$ data, nuclear modification of parton distributions 
and $p_T$ broadening give about 10-30\% increase in the $p_T$ 
spectra in central $Au+Au$ collisions relative to $pp$ without
parton energy loss, as shown in Fig.~\ref{fig:r-rhic}.

Many recent theoretical studies \cite{loss} predict a large amount
of energy loss by a propagating energetic parton in a dense partonic
medium. In $AA$ collisions, we model the effect of parton
energy loss by the modification of parton fragmentation functions \cite{whs}.
This will lead to suppression of large $p_T$ hadrons for a non-vanishing
parton energy loss relative to the null scenario.
In terms of the nuclear modification factor, $R_{AB}(p_T)$ will become 
smaller than 1 at large $p_T$ due to parton energy loss as shown 
in Fig.~\ref{fig:r-rhic}. It is also shown that the $p_T$ 
dependence of $R_{AB}(p_T)$ is sensitive to the energy dependence of 
the parton energy loss. For more accurate determination of the
energy dependence of $dE/dx$ one should measure the high $p_T$ hadron
suppression in direct $\gamma$-tagged events \cite{whs}.

Preliminary RHIC data \cite{phenix,star} 
have indeed shown evidence of suppression of large $p_T$ hadron spectra
for the first time. This is extraordinary, given that no suppression
is seen at SPS \cite{xnw00}.
Shown in Fig.~\ref{fig:pi0} are $\pi_0$ spectra \cite{phenix}
in central $Au+Au$ collisions as compared to pQCD parton model 
calculation. The model result for $p\bar p$ collisions 
at $\sqrt{s}=200$ GeV agrees well with the UA1 data. Both the calculation
and data for $Au+Au$ collisions are scaled by 
$\langle N_{\rm part}\rangle=\int d^2b T_{AB}(b)$ for the given centrality
in order to compare to $pp(\bar p)$ results. The preliminary data clearly
show suppression of high $p_T$ hadron spectra relative to $pp$.
The data are consistent with the parton model calculation
with an average $dE/dx\approx 0.25$ GeV/fm. I want to emphasize that
this is only a phenomenological value averaged over the entire evolution
history of the dense matter. It does not reveal the distance and density
dependence of the $dE/dx$. Since the initial parton density decreases
very fast due to longitudinal expansion, such a small average $dE/dx$
still implies very large parton energy loss in the very early stage
of the dense system. Future analysis of the energy and centrality 
dependence \cite{gvw} of the high $p_T$ hadron suppression is important to
understand the consistency between CERN-SPS and RHIC results and
search for the critical initial parton density at which jet quenching
becomes significant in heavy-ion collisions.

In non-central $AA$ collisions, the total parton propagation length should
depend on the azimuthal direction. Therefore, parton energy loss can 
also cause azimuthal anisotropy of hadron spectra at large $p_T$
in non-central $AA$ collisions \cite{gvw}. Such azimuthal 
anisotropy can be calculated in the same parton model with a given
parton energy loss. Shown in Fig.~\ref{fig:v2} is the calculated \cite{gvw}
$v_2(p_T)$ as a function of $p_T$ as compared to STAR preliminary 
data \cite{star}. The hadron spectra at low $p_T<2$ GeV/$c$
are assumed to be given by the hydrodynamic model and $v_2$ increases
linearly with $p_T$. However, at $p_T>2$ GeV/$c$, hydrodynamics will 
fail and $v_2$ should be determined by the dynamics of parton 
propagation in the dense medium. The parton model gives very different $p_T$
dependence. The point where $v_2(p_T)$ deviates significantly from the hydro 
model signals the onset of the contribution of hard processes. As shown in
the figure, $v_2$ at large $p_T$ is very sensitive to $dE/dx$ or the
initial parton density. It is an alternative measurement of parton
energy loss or the initial parton density. Since suppression of hadron 
spectra and non-vanishing $v_2$ at large $p_T$ are two consequences of 
parton energy loss, one should be able to explain both experimental 
measurements and their centrality dependence with the same
given parton initial density.

\section{Summary}
I have discussed the importance of multiple scattering,
coherence and
the interplay between soft and hard processes in the modeling of
high-energy heavy-ion collisions. Both the nuclear dependences of 
rapidity density and $p_T$ spectra have shown the onset of hard processes.
The preliminary RHIC data show evidence of high $p_T$ hadron
suppression due to parton energy loss. I argued that the combination of
$v_2$ and hadron spectra suppression will help us to determine
the initial gluon density in heavy-ion collisions at RHIC.

\section*{Acknowledgement}
Many results reported here are from work in collaboration with
M. Gyulassy, I. Vitev and E. Wang. I also thank G. David, A. Drees, B. Jacak,
P. Jacobs, F. Messer and R. Snellings for discussions.

\end{document}